\documentclass[prl,twocolumn,amsmath,amssymb,superscriptaddress,showpacs]{revtex4}
\usepackage{graphicx}
\usepackage{float}
\usepackage{bm}

\begin{document}

\title{Comment on ``Pronounced Enhancement of the Lower Critical Field and Critical Current Deep in the Superconducting State of PrOs$_{\bm{4}}$Sb$_{\bm{12}}$''}

\author{D.~E. MacLaughlin}
\affiliation{Dept.\ of Physics \& Astronomy, Univ.\ of California, Riverside, California 92521, USA}
\author{A.~D. Hillier}
\affiliation{ISIS, Rutherford Appleton Laboratory, Chilton, Didcot, OC11 0QX, UK}
\author{J.~M. Mackie}
\affiliation{Dept.\ of Physics \& Astronomy, Univ.\ of California, Riverside, California 92521, USA}
\author{Lei Shu}
\affiliation{Dept.\ of Physics \& Astronomy, Univ.\ of California, Riverside, California 92521, USA}
\affiliation{Dept.\ of Physics \& IPAPS, Univ.\ of California, San Diego, La Jolla, California 92093, USA}
\author{Y. Aoki}
\author{D. Kikuchi}
\author{H. Sato}
\author{Y. Tunashima}
\affiliation{Dept.\ of Physics, Tokyo Metropolitan Univ.\ Tokyo 192-0397, Japan}
\author{H. Sugawara}
\affiliation{Fac.\ of Integrated Arts \& Sciences, Univ.\ of Tokushima, Tokushima 770-8502, Japan}

\date{January 18, 2010}


\maketitle

Cichorek \textit{et al.}~\cite{CMSF05} reported enhancements of the lower critical field~$H_{c1}(T)$ and critical current~$I_c(T)$ in superconducting PrOs$_{4}$Sb$_{12}$ below a transition temperature~$T_{c3} \approx 0.6$~K, and speculated that this reflects a transition between superconducting phases. Features have been observed near $T_{c3}$ in other properties, but not in the specific heat. We report muon spin rotation ($\mu$SR) measurements of the penetration depth $\lambda$ in the vortex state of PrOs$_{4}$Sb$_{12}$ near $H_{c1}$, that to high accuracy exhibit no anomaly $T_{c3}$ and therefore cast doubt on the putative phase transition. 

In a Type-II superconductor $H_{c1} = \Phi_0(\ln\kappa + c)/4\pi\lambda^2$, $c \approx 0.5$, where $\Phi_0$ is the flux quantum and $\kappa$ is the Ginz\-burg-Landau parameter~\cite{Bran03}. Modification of the superfluid density $\rho_s$ by a phase transition should affect both $\lambda = (mc^2/4\pi e^2\rho_s)^{1/2}$ and $H_{c1}$. A feature observed near $T_{c3}$ in rf inductive measurements of $\lambda$~\cite{CSSS03} is too small to account for the observed enhancement of $H_{c1}$~\cite{CMSF05}.

In the transverse-field $\mu$SR technique the spectrum of muon precession frequencies gives the local-field distribution function, which depends on $\lambda$ in the vortex state~\cite{Bran03}. $\mu$SR experiments were carried out at the ISIS $\mu$SR facility on a polycrystalline sample of PrOs$_4$Sb$_{12}$. Strong de Haas-van Alphen signals obtained from similarly-prepared crystals~\cite{ATKS03} attest to their high quality. Data were taken in low applied fields~$H = 25$ and 40~Oe, the former corresponding to an internal field~$H' \sim 32$~Oe at $H' = H_{c1}(T)$ after demagnetization correction. This is close to the estimated unenhanced value of $H_{c1}(0)$~\cite{CMSF05}, so that in the absence of enhancement $H'$ should be $> H_{c1}$. 

\begin{figure}[ht]
\begin{center}
\includegraphics*[clip=,width=0.95\columnwidth]{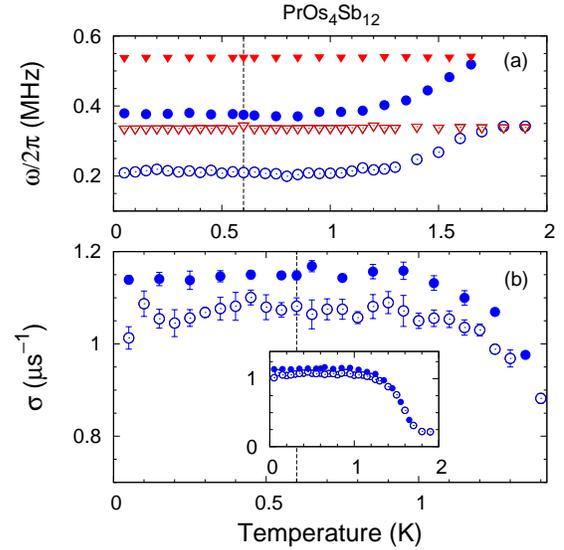}
 \vspace{-90pt} \caption{(Color online) $\mu$SR in superconducting PrOs$_{4}$Sb$_{12}$ at applied fields $H = 25$~Oe (open symbols) and 40~Oe (filled symbols). Dashed lines: $T_{c3}$ from Ref.~\protect\cite{CMSF05}. (a)~Muon spin precession frequency $\omega_\mu(T)$. Circles: sample. Triangles: reference. (b)~Gaussian muon spin relaxation rate $\sigma(T)$.}
\label{fig:POS25_frq-rlx}
\end{center}
\vspace{-30pt}
\end{figure}

The data are well fit by the Gaussian relaxation function~$G(t) = \exp(-\frac{1}{2}\sigma^2 t^2)\cos(\omega_\mu t + \theta)$. Figure~\ref{fig:POS25_frq-rlx} gives the average muon spin precession frequency~$\omega_\mu(T)$ and relaxation rate~$\sigma(T)$ at 25 and 40~Oe.
There is no discernible anomaly at $T_{c3}$, and no evidence that $H' < H_{c1}$. Similar results are found at higher fields~\cite{MSHS08}.

The rms width~$\delta B_{\rm rms}$ of the field distribution in the vortex state is estimated by $\sigma/\gamma_\mu$, where $\gamma_\mu$ is the muon gyromagnetic ratio. In the London model $\delta B_{\rm rms}^2 = 0.00371 \Phi_0^2/\lambda^4$~\cite{Bran03}, i.e., $H_{c1}$ and $\delta B_{\rm rms}$ are (essentially) proportional to $\Phi_0/\lambda^2$ and therefore to each other: $H_{c1}/\delta B_{\rm rms} = 1.31 (\ln\kappa + c) \approx 5.0$ in PrOs$_{4}$Sb$_{12}$. Thus a $\sim$50\% enhancement of $H_{c1}(0)$~\cite{CMSF05} implies a similar enhancement of $\sigma(0)$ contrary to our results. From $\sigma(0)$ we estimate~$H_{c1}(0) \approx 50$~Oe, of the order of the observed value, so that other broadening mechanisms, such as vortex-lattice disorder or a distribution of demagnetizing fields, are unlikely to dominate the muon relaxation. 

These results and the absence of a specific heat anomaly at $T_{c3}$ are evidence against a phase transition associated with the enhancement of $H_{c1}$ at low temperatures in PrOs$_{4}$Sb$_{12}$. The enhanced critical current suggests that flux pinning effects, which generally become stronger at low temperatures, are involved.

Work supported by the U.S. NSF, Grant no.~0422674 (Riverside), and a Grant-in-Aid for Scientific Research on Priority Areas (19052003) (Tokyo).

\end{document}